\documentclass[showpacs,showkeys, amsmath,amssymb, preprint]{revtex4-1}
\usepackage{graphicx}
\usepackage{epsfig}
\usepackage{bm}

\usepackage{fontenc}
\usepackage{graphicx}
\usepackage[dvips]{hyperref}
\usepackage{bm}
\usepackage{amsmath}
\def\beq{\begin{eqnarray}}

\def\eeq{\end{eqnarray}}

\begin{document}

\title[Scalar perturbations of two-dimensional Horava-Lifshitz Black Holes]{Scalar Perturbations of two-dimensional Horava-Lifshitz Black Holes}

\author{Miguel Cruz}
\email{miguel.cruz@ucv.cl}
\address{Instituto de F\'\i sica, Pontificia Universidad Cat\'olica de Valpara\'\i so, Casilla 4950, Valpara\'\i so}

\author{Manuel Gonzalez-Espinoza}
\email{manuelgonza1985@gmail.com}
\address{Instituto de F\'\i sica, Pontificia Universidad Cat\'olica de Valpara\'\i so, Casilla 4950, Valpara\'\i so}

\author{Joel Saavedra}
\email{joel.saavedra@ucv.cl}
\address{Instituto de F\'\i sica, Pontificia Universidad Cat\'olica de Valpara\'\i so, Casilla 4950, Valpara\'\i so}

\author{Diego Vargas-Arancibia}
\email{vargas7042@gmail.com}
\address{Instituto de F\'\i sica, Pontificia Universidad Cat\'olica de Valpara\'\i so, Casilla 4950, Valpara\'\i so}


\begin{abstract}
In this article, we study the stability of black hole solutions found in the context of dilatonic Horava-Lifshitz gravity in $1+1$ 
dimensions by means of the quasinormal modes approach. In order to find the corresponding quasinormal modes, we consider the perturbations of massive 
and massless scalar fields minimally coupled to gravity. In both cases, we found that the quasinormal modes have a discrete spectrum 
and are completely imaginary, which leads to damping modes. For a massive scalar field and a non-vanishing cosmological constant, our 
results suggest unstable behaviour for large values of the scalar field mass.     
\end{abstract}


\pacs{04.70.-s, 04.70.Bw, 04.50.Kd}

\maketitle

\date{\today}

\section{Introduction}

For quite some time, physicists have considered Einstein's General Relativity (GR) to be an effective theory of gravity. Therefore, in order 
to find the happy marriage between quantum theory and gravity, we need to know the underlying fundamental theory of gravity. One recent 
proposal in this address has been the Horava-Lifshitz (HL) theory \cite{Horava:2008jf}, that is, a power-counting renormalizable theory 
with consistent ultraviolet (UV) behaviour. Furthermore, the theory has one fixed point in the infrared (IR) limit namely 
GR \cite{Horava:2008jf, Horava:2008ih, Horava:2009if, Horava:2009uw}. In terms of the above, black holes (BH) are important 
solutions for field equations in any gravitational theory, including those such as Einstein-Hilbert, Brans-Dicke, HL, $f(R)$, 
string theories and any generalisation or modification of Einstein's gravity. At the quantum level, BH play the same role as hydrogen atom and we hope they 
give us some clues about the observables of any quantum theory of gravity. As such, it is important to study the physical properties of 
BH solutions, such as decay rate, greybody factors, or their quasinormal modes. Quasinormal modes (QNMs), known as ``ringing'' in 
BH, are very important in order to understand the classical and quantum aspects of BH physics. The QNMs give us hints about the 
stability of BH under consideration, as in this paper, and can be used to compute the spectrum of the area operator using the semiclassical approach 
developed by Hod \cite{Hod:1998vk}. The determination of QNMs is based on the dynamics of matter fields and on the metric perturbations 
in the BH background. In this work indeed, we are interested in the stability of the $1+1$-dilatonic HL BH using a QNMs' approach; QNMs 
associated with the perturbations of different fields have been considered in different works \cite{Kokkotas:1999bd}, including those
involving dS and AdS space \cite{Horowitz:1999jd,c2,c3,c4,c5,Chan:1999sc,Wang:2001tk,Konoplya:2003dd} and higher dimensional models, 
where the QNMs can be computed for a brane situated in the vicinity of a $D$-dimensional BH \cite{Konoplya2}. A similar situation occurs in $2+1$ dimensions
\cite{Chan:1996yk,c44,Crisostomo:2004hj}, and for acoustic BH \cite{c1, Lepe:2004kv,Saavedra:2005ug}. QNMs of dilatonic BH in $3+1$ 
dimensions can be found in Refs. \cite{Ferrari:2000ep, Konoplya:2002ky,Fernando:2003wc}. Two-dimensional theories of gravity have 
recently attracted much attention\cite{Robinson:2005pd,Myung:2000hk,Torii:1998gm} as simple toy models that possess many features of 
gravity in higher dimensions. They also have BH solutions which play important roles in revealing various aspects of spacetime 
geometry and quantization of gravity, and are also related to string theory \cite{Teo:1998kp,McGuigan:1991qp}. The QNMs of 
$1+1$ dilatonic BH for scalar and fermionic perturbations were studied in 
\cite{Becar:2007hu, Becar:2010zz, LopezOrtega:2011sc, Becar:2014jia}.

The determination of QNMs for a specific geometry implies solving the field equations for different types of perturbations 
(scalar, fermionic, vectorial, etc.), with suitable boundary conditions that reflect the fact that this geometry describes a BH. 
The QNMs of a classical scalar perturbation of BH are defined as the solutions of the Klein-Gordon equation characterised by purely
ingoing waves at the horizon, $\Phi \sim e^{-i\omega (t+r)}$, since, at least classically, an outgoing flux is not allowed at the
horizon. In addition, one has to impose boundary conditions on the solutions in the asymptotic region (infinity), and for that reason it is
crucial to use asymptotic geometry for the spacetime under study. In the case of an asymptotically flat spacetime, the condition 
we need to impose over the wave function is to have a purely outgoing wave function $\Phi \sim e^{-i\omega (t-r)}$ at the infinity 
\cite{Horowitz:1999jd}. In general, the QNMs are given by $\omega _{QNM}=\omega _{R}+i\omega _{I}$, where $\omega _{R}$ and
$\omega _{I}$ are the real and imaginary parts of the frequency $\omega _{QNM}$, respectively. Therefore, the study of QNMs can be
implemented as one possible simple alternative test for studying the stability of the system. In this sense, any imaginary frequency with the wrong sign
would mean an exponentially growing mode, rather than a damping one.

The organisation of this article is as follows: In Section II, we describe briefly the HL theory and specify the $1+1$-dilatonic BH 
solutions. In Section III, we compute the QNMs and explore the criteria for the stability of the two BH metrics under consideration. We 
finish with conclusions in Section IV.

\section{Generalities of the Horava-Lifshitz gravity}
In the following, we will describe the HL theory as developed in Ref. \cite{Horava:2009uw}. The HL theory provided a new 
approach to quantum gravity and its principal idea is based on the breaking of the Lorentz invariance by equipping the 
spacetime with additional geometric structure, a prefered foliation which defines the splitting of the coordinates into 
space and time; in this theory the Lorentz invariance is assumed to appear only at the low energies limit. One can decompose 
the spacetime as follows
\begin{equation}
ds^{2} = (N^{2}-N_{i}N^{i})dt^{2}-2N_{i}dx^{i}dt-h_{ij}dx^{i}dx^{j}, 
\end{equation}
where $N, N^{i}$ are the lapse and shift functions respectively and $h_{ij}$ is the three-dimensional metric. The action is given by
\begin{equation}
S = \frac{M^{2}_{Pl}}{2}\int d^{3}xdt\sqrt{h}N\left(K_{ij}K^{ij}-\gamma K^{2}-\mathcal{V} \right), 
\label{eq:HLaction}
\end{equation}
where $M_{Pl}$ is the Planck mass, $\gamma$ is a dimensionless constant and $K_{ij}$ is the well-known extrinsic curvature tensor, 
which is stated in the ADM formulation as
\begin{equation}
K_{ij}= \frac{1}{2N}\left(\dot{h}_{ij}-\nabla_{i}N_{j}-\nabla_{j}N_{i} \right), 
\end{equation}
being $K$ its trace. The last term in (\ref{eq:HLaction}), $\mathcal{V}$, is invariant under three-dimensional diffeomorphisms and is 
known as the ``potential'' term. This term is a function of the three-dimensional metric and its derivatives. In explicit form we have
\begin{equation}
\mathcal{V} = -\xi R + \frac{1}{M^{2}_{Pl}}\left(\pi_{1}\Delta R + \pi_{2}R_{ij}R^{ij} + ... \right) +
\frac{1}{M^{4}_{Pl}}\left(\sigma_{1}\Delta^{2} R + \sigma_{2}R_{ij}R^{jk}R^{i}{}_{k} + ... \right),  
\end{equation}
where $\xi, \pi_{n}, \sigma_{n}$ are coupling constants, $R_{ij}$ and $R$ are the Ricci tensor and the scalar curvature constructed 
with the spatial metric. $\Delta:= h^{ij}\nabla_{i}\nabla_{j}$. The introduction of the ``potential'' term in (\ref{eq:HLaction}) 
improve the UV behavior of the graviton propagator and additionally leads to different scaling of space and time
\begin{equation}
\mathbf{x} \rightarrow \rho^{-1}\mathbf{x}, \ \ t \rightarrow \rho^{-3}t, \ \ N \rightarrow N, \ \ N^{i} \rightarrow \rho^{2}N_{i}, 
\ \ h_{ij}\rightarrow h_{ij}. 
\end{equation}
When the lapse function depend only on time, $N=N(t)$, we say that we are dealing with the ``projectable'' version of the HL theory 
and the ``non-projectable'' version is given when the lapse function can depend on space and time. In Ref. \cite{Blas} an extension 
of the non-projectable version of HL gravity was made by the introduction of a extra mode in the ``potential'' term, i.e., 
$\mathcal{V}(h_{ij})\rightarrow \mathcal{V}(h_{ij},a_{i})$. It was shown that this extra mode can acquire a regular quadratic 
Lagrangian. The extra mode is given by
\begin{equation}
a_{i} := \frac{\partial_{i}N}{N}. 
\end{equation}
Geometrically this vector represents the proper acceleration of the unit normals to the spatial slices. For the two-dimensional case 
there are only two terms that contribute to the quadratic Lagrangian: $R$ and $a_{i}a^{i}$.

\subsection{Lowest dimensional Horava-Lifshitz Black Hole}
\label{sec:HLbh}
The HL gravity has two-dimensional solutions that characterise dilatonic BH and can be used to study the physical properties of BH in 
general; furthermore, some features of this theory, due to the fact that it utilizes two dimensions, open the possibility of 
understanding physical consequences in higher dimensional theories. As a summary, let us start with the HL-dilaton gravity in 
two dimensions presented in Ref. \cite{HL}, 
\begin{equation}
S = S_{HL}+S_{\phi},
\end{equation}
where, as mentioned before, the quadratic Lagrangian for the HL theory in two dimensions comes from the contribution of the terms 
$R$ and $a_{i}a^{i}$
\begin{equation}
S_{HL} = \frac{M^{2}_{Pl}}{2}\int dtdx\sqrt{g}\left((1-\lambda)K^{2}+\eta g^{11}a_{1}a_{1} \right), 
\end{equation}
and 
\begin{equation}
S_{\phi} = \int dtdxN\sqrt{g}\left[\frac{1}{2N}\left(\partial_{t}\phi -N^{1}\nabla_{1}\right)^{2}-\alpha(\nabla_{1}\phi)^{2}-V(\phi) 
-\beta \phi \nabla^{1}a_{1}-\varsigma \phi a^{1}\nabla_{1}\phi \right], 
\end{equation}
where $\alpha$, $\beta, \eta$ and $\varsigma$ are constants. Using the fact that $K=0$ and admitting $N_{1} = 0$ together with the relativistic
limit $\beta=\varsigma=0$, we are left with the action
\begin{equation}
S= \frac{M^2_{Pl}}{2}\int dtdx \left (-\frac{1}{2}\eta N^2a^2_1+\alpha N^2\phi'^2-V(\phi)\right).
\label{eq:action} 
\end{equation}
From now on, the prime denotes derivative with respect the coordinate $x$. In two dimensions the extra mode $a_{i}$ 
is simply $a_1=\partial_{1}\ln N = \left(\ln N \right)'$. In Ref. \cite{HL}, a new set of BH solutions in two-dimensional HL gravity 
was found for action (\ref{eq:action}). The solutions are described by
\begin{equation}
N(x)=\sqrt{\frac{A}{\eta}x^2-2C_1x+\frac{B}{\eta x}+\frac{C}{3 \eta x^2}+2C_2}\,\,\,\,,
\label{eq:solutionN} 
\end{equation}
and 
\begin{equation}
\phi(x)=\ln \sqrt{\frac{A}{\eta}x^2-2C_1x+\frac{B}{\eta x}+\frac{C}{3 \eta x^2}+2C_2}\,\,\,\,,
\label{eq:solutionphi} 
\end{equation}
these solutions were obtained by using the quantity
\begin{equation}
V_{\phi}(x) = A + \frac{B}{x^{3}}+\frac{C}{x^{4}}, 
\end{equation}
the derivative of the scalar potential given as a function of an implicit scalar field which in turns depends on the spatial 
coordinate. This was done because for generalized potentials is not always possible obtain analytical solutions. We would like to 
focus our attention on the following two cases,

$\bullet$  First case: Described by fixing the constants in the following way; $A=B=C=0$, $C_1= -M$, 
$C_2=-1/2$ and $\eta =1$. In this case $V_{\phi}=0$. Therefore, the metric for this solution can be written as follows
\begin{equation}
ds^{2} = -(2Mx -1)dt^{2} + \frac{1}{2Mx - 1}dx^{2},
\label{eq:lowest} 
\end{equation}
where the parameter $M$ is related to the lapse function $N$. This solution was found for the first time in \cite{Mann:1991md}.

$\bullet$ Second Case: Here we fix the constants as $A=\Lambda$, $B=C=0$, $C_1=-M$ and $C_2=-\frac{\epsilon }{2}$; therefore, we 
have $V_{\phi}=\Lambda$, and the solution is given by
\begin{equation}
ds^{2} = \left[\left(\frac{\Lambda}{\eta}\right)x^{2} + 2Mx - \epsilon \right]dt^{2} + 
\frac{1}{\left(\frac{\Lambda}{\eta}\right)x^{2} + 2Mx - \epsilon}dx^{2}.
\label{eq:sol2}
\end{equation} 
The horizon of the black hole is located at
\begin{equation}
x_{\pm} = -\frac{\eta M}{\Lambda} \pm \sqrt{\frac{\eta}{\Lambda}\left(\frac{\eta M^{2}}{\Lambda} + \epsilon \right)}.
\label{eq:sol1}
\end{equation}
If we define the variables $u = \sqrt{\Lambda / \eta}x + \sqrt{\eta / \Lambda}M$ and 
$u_{+} = \sqrt{(\eta / \Lambda)M^{2} + \epsilon}$ one gets
\begin{equation}
ds^{2} = -\left(u^{2} - u_{+}^{2} \right)dt^{2} + 
\frac{l^{2}}{\left(u^{2} - u_{+}^{2}\right)}du^{2},
\label{eq:sol3}
\end{equation}
which is a suitable expression to study the quasinormal modes of this black hole, and we have defined $l = (\Lambda/\eta)^{1/4}$. In 
this new coordinate system, the horizon of the  black hole is located at $u = u_{+}$. The spacetimes, described by (\ref{eq:lowest}) 
and (\ref{eq:sol2}), are conformally flat \cite{Mann}.

\section{Quasinormal modes}
\label{sec:qnms}
In order to study the QNMs, we consider a scalar field minimally coupled to gravity propagating in the background of 
the two-dimensional HL BH. We consider the following action for the scalar field
\begin{equation}
S[\varphi] = \int d^{2}x \sqrt{-g}\left(-\frac{1}{2}(\nabla \varphi)^{2} - \frac{1}{2}m^{2}\varphi^{2} \right),
\end{equation}
where $m$ is the mass of the scalar field. From the variation of $\delta \phi$ the field equation is given by
\begin{equation}
 \Box \varphi - m^{2}\varphi = 0,
\label{eq:eom}
\end{equation}
where $\Box$ is the D'Alambertian operator, in the following sections, we will solve the Klein-Gordon equation (\ref{eq:eom}) for the spacetimes described in the previous section.

\subsection{Spacetime metric $ds^{2} = -(2Mx -1)dt^{2} + \frac{1}{2Mx - 1}dx^{2}$ }

\subsubsection{Massive Scalar Field}
The case of a massive scalar field perturbing the background described by the metric (\ref{eq:lowest})
was discussed for first time in \cite{Mann, Mann:1991md}. If we use $\varphi(t,x)=e^{-i\omega t}\varphi(x)$ the equation of motion (\ref{eq:eom}), 
is represented by
\begin{equation}
(2Mx-1)\frac{d^{2}\varphi(x)}{dx^{2}} + 2M \frac{d\varphi(x)}{dx} + \frac{\omega^{2}}{2Mx-1}\varphi(x)-m^2\varphi(x)= 0.
\label{eq:nobessel0}
\end{equation}
For this metric and massive scalar field, the QNMs were computed in \cite{Lopez}, where the authors claimed that the solution for 
the QNMs are completely different from the standard case where the QNMs have a discrete spectrum. They proposed a real and continuous 
spectrum for the QNMs of a scalar perturbation. In the next section, we look at this situation, but instead use the confluent hypergeometric function 
$_{0}F_1(a,b;x)$ place of the modified Bessel functions. Now, using the tortoise coordinate defined by $x_{*}=\frac{1}{2M}\ln(2Mx-1)$ 
we can write the Eq. (\ref{eq:nobessel0}) as a Schroedinger type equation with effective potential \cite{Lopez}
\begin{equation}
V_{eff} =m^{2}e^{2Mx_{*}},
\label{eq:potential}  
\end{equation}
this potential diverges when $x_{*}\rightarrow \infty$. If we consider the variable $z=\frac{m}{M}e^{Mx_{*}}$ \cite{Mann:1991md}, 
after some straightforward algebra, the equation of motion (\ref{eq:nobessel0}) can be written as the Bessel equation,
\begin{equation}
z^2\frac{d^2\varphi(z)}{dz^2}+z\frac{d\varphi(z)}{dz}+\left(\nu^2+z^2\right)\varphi(z)=0,
\label{eq:bessel}
\end{equation}
where $\nu=\frac{i\omega}{M}$; this equation can be transformed into the confluent hypergeometric equation using the 
change $\varphi(z)=(2iz)^{\nu}e^{-2z}F(z)$ \cite{Libro}, and we find
\begin{equation}
z\frac{d^2F(z)}{dz^2} + (2\nu+1-2iz)\frac{dF(z)}{dz}+2i\left(\nu+\frac{1}{2}\right)F(z)=0,
\label{eq:kumer}
\end{equation}
whose solution is given in terms of the confluent hypergeometric, or Kummer, functions
\begin{equation}
F(z)=A\Phi(\nu+\frac{1}{2},2\nu+1;2iz)+B(2iz)^{-2\nu}\Phi(-\nu+\frac{1}{2},1-2\nu;2iz),
\label{eq:kumersolgene}
\end{equation}
where $A,B$ are constants and the number $\nu$ in general need not be an integer \cite{Libro}. In the following we will 
consider two cases for the $\nu$ parameter since we are interested in exploring all its possible values.\\

$\bullet$ non-integer $\nu$ 
 
In order to compute the QNMs, we need to impose adequate boundary conditions that represent a purely outgoing wave at 
infinity and purely ingoing wave near the horizon of the BH (a condition often used in flat spacetime). Another situation occurs 
when the asymptotic behaviour of the spacetime is not flat, e.g. asymptotically AdS space; in these kind of spaces, the potential 
diverges at infinity, and we can therefore impose $\varphi=0$ (Dirichlet boundary condition) or $\frac{d\varphi}{dx}=0$  
(Neumann boundary condition) at infinity. As we can see from Eq. (\ref{eq:potential}), in this case we have an asymptotically AdS space. 
Therefore, we need to apply boundary conditions to the QNMs over the general solution of Eq. (\ref{eq:kumer}), which is given by
\begin{equation}
\varphi(z)=Ae^{-2z}z^{\nu}\Phi(\nu+\frac{1}{2},2\nu+1;2iz)+Be^{-2z}z^{-\nu}\Phi(-\nu+\frac{1}{2},1-2\nu;2iz),
\label{eq:kumersolgen}
\end{equation}
to satisfy the boundary conditions properly, we set $A=0$ in order to have only ingoing waves at the horizon ($z=0$). 
The asymptotic behavior of $\Phi$ at infinity is given by \cite{energy}
\begin{equation}
\Phi(p,q;z)\rightarrow\frac{\Gamma(q)}{\Gamma(p)}z^{p-q}e^z,
\label{eq:kumerlimit}
\end{equation}
and therefore our solution at the infinity ($z \rightarrow \infty$) reads as follow,
\begin{equation}
\varphi(z)\sim Bz^{-\frac{1}{2}}e^{2(i-1)z}\frac{\Gamma(1-2\nu)}{\Gamma(-\nu+1/2)}.
\label{eq:sollimit}
\end{equation}
We can see that the scalar field vanishes as $z \rightarrow \infty$, this confirms the absence of QNMs for this HL BH under 
scalar perturbations, a similar situation was found in Ref. \cite{Crisostomo:2004hj} for the QNMs of the extremal BTZ BH. 
The conclusion of this case was discussed in \cite{Lopez}, and represents a continuous spectrum. Additionaly, if 
we impose the Neumann boundary condition, we obtain a similar asymptotic vanishing behavior for the flux
\begin{equation}
J_r(z)=\varphi^*(z)\frac{d\varphi(z)}{dz}-\varphi(z)\frac{d\varphi^*(z)}{dz}.
\label{eq:current}
\end{equation}
In light of the meaning of QNMs, for any BH perturbation its geometry produces damped oscillations, this is 
the so-called ringing in BH. It is well known that the frequencies of these oscillations and their damping periods are completely 
fixed by the BH properties, and as such, are independent of the nature of the initial perturbation. In Ref. \cite{Lopez} it was shown 
that, for scalar perturbations the oscillations have a continuous spectrum and are not discrete, as would be 
expected for a BH. This result is very strange and in our opinion, devoid of physical meaning; this is because it is well known that oscillations of 
QNMs are similar to normal modes of a closed system. In the next section, we consider the second solution of the confluent hypergeometric 
equation and show that QNM oscillations have a discrete spectrum.

$\bullet$ $\nu$ integer

Now we are considering the case where $\nu$ is an integer number. For $\nu$ integer, the solution of the Eq. (\ref{eq:kumer}) changes. 
In the expression (\ref{eq:kumersolgen}), the function $\Phi(p,q;z)$ of the second term must be replaced by \cite{macdonal}
\begin{eqnarray}
W(\alpha,\gamma;z)&=&M(\alpha,\gamma;z)\left(\ln z+\psi(1-\alpha)-\psi(\gamma)+C\right)+\sum_{n=1}^{\infty}\frac{\Gamma(n+\alpha)\Gamma(\gamma)B_nz^n}{\Gamma(\alpha)\Gamma(n+\gamma)n!}+\nonumber \\  \nonumber
&+&(-1)^{\gamma}\sum_{n=1}^{\infty}\frac{\Gamma(\gamma)\Gamma{n+\alpha-\gamma+1}\Gamma(\gamma-n-1)(-1)^n}{\Gamma(\alpha)n!z^{\gamma-n-1}}\label{current}\,,
\end{eqnarray}
where 
\begin{equation}
\psi(\alpha)=\frac{\Gamma'(\alpha)}{\Gamma(\alpha)},
\label{eq:current}
\end{equation}
represents the digamma function, $C=0.577216...$ is the Euler's constant, and 
\begin{equation}
B_n=\left(\frac{1}{\alpha}+\frac{1}{\alpha+1}+...+\frac{1}{\alpha+n-1}\right)-\left(\frac{1}{\gamma}+\frac{1}{\gamma+1}+...+\frac{1}{\gamma+n-1}\right).
\label{eq:currt}
\end{equation}
Then, we have
\begin{equation}
\varphi(z)=Ae^{-2z}z^{\nu}\Phi(\nu+\frac{1}{2},2\nu+1;2iz)+Be^{-2z}z^{-\nu}W(p,q;z).
\label{eq:kumersolgen6}
\end{equation}
In order to have only ingoing waves at the horizon ($z=0$), we set $A=0$. The asymptotic behaviour of the $W$-function at infinity 
is given by
\begin{equation}
W(p,q;y)\rightarrow  \pi \cot(\pi p)\frac{\Gamma(q)}{\Gamma(p)}y^{p-q}e^y.
\label{eq:W}
\end{equation}
Therefore, the general solution at infinity reads as follows,
\begin{equation}
\varphi(z)\sim B\pi \cot\left[-\pi \left(\nu+\frac{1}{2}\right)\right]z^{-\left(4\nu+\frac{3}{2}\right)}\frac{\Gamma(2\nu+1)}{\Gamma(-\nu-1/2)}
e^{2(i-1)z}.
\label{eq:sollimitW}
\end{equation}
If we consider that $\nu + 1/2=\frac{2n+1}{2}$, where $n$ is an integer number, we are able to fulfill the Dirichlet boundary 
condition at infinity. From this result we can obtain the frequency of the QNMs as
\begin{equation}
\omega=-inM,
\label{eq:omegaW}
\end{equation}
and using the Neumann condition for a vanishing flux at infinity, we obtain the same result for the QNMs as expressed 
in Eq. (\ref{eq:omegaW}).

\subsubsection{Massless Case}

For the metric (\ref{eq:lowest}), when $m=0$, we use the standard definition for QNMs, the Klein-Gordon equation (\ref{eq:eom}) 
which reads 
\begin{equation}
(2Mx-1)\frac{d^{2}\varphi(x)}{dx^{2}} + 2M \frac{d\varphi(x)}{dx} + \frac{\omega^{2}}{2Mx-1}\varphi(x) = 0,
\label{eq:nobessel}
\end{equation}
where we have assumed $\phi(t,x) = \varphi(x)e^{-i\omega t}$. If we define the quantity $x_{+} = 1/2M$ and the change of variable 
$z = 1 - x_{+}/x$ we can write equation (\ref{eq:nobessel}) as follows
\begin{equation}
(1-z)^{2}\frac{d^{2}\varphi}{dz^{2}}-2(1-z)\frac{d\varphi}{dz}+\frac{(1-z)}{z}\frac{d\varphi}{dz}+\left(\frac{\tilde{\omega}}{z}\right)^{2}
\varphi = 0, 
\end{equation}
where $\tilde{\omega} = x_{+}\omega$. Note that in the new coordinate $z$, the horizon of the BH is located at $z = 0$ and infinity at 
$z=1$. With the change $\varphi(z) = z^{\alpha}(1-z)^{\beta}F(z)$, the last equation reduces to the hypergeometric differential 
equation for the function $F(z)$, that is,
\begin{equation}
z(1-z)F''(z) + (c-(a+b+1)z)F'(z)-abF(z)=0.
\label{eq:hyper} 
\end{equation}
In this case the coefficients $a$, $b$ and $c$ are given by the relations
\begin{eqnarray}
c &=& 2\alpha +1, \\
a+b &=& 2\alpha + 2\beta + 1, \\
ab &=& \alpha(\alpha - 1) + \beta(\beta-1)+2\alpha + 2\beta + 2\alpha \beta,
\end{eqnarray}
providing the expressions for the coefficients
\begin{eqnarray}
a &=& \alpha +\beta,\\
b &=& 1+\alpha +\beta,
\end{eqnarray}
and for the exponents we obtain
\begin{equation}
 \alpha = \beta = -ix_{+}\omega.
\end{equation}
Without loss of generality, we have chosen the negative signs for the exponents. The solution of the radial equation reads
\begin{equation}
 F(z) = C_{1}F_{1}(a,b,c;z)+C_{2}z^{1-c}F_{1}(a-c+1,b-c+1,2-c;z),
\end{equation}
where $C_{1}$ and $C_{2}$ are arbitrary constants and $F_{1}(a,b,c;z)$ is the hypergeometric function. The solution for $\varphi(z)$
is given by
\begin{equation}
 \varphi(z) = C_{1}z^{-ix_{+}\omega}(1-z)^{-ix_{+}\omega}F_{1}(a,b,c;z) + C_{2}z^{ix_{+}\omega}(1-z)^{-ix_{+}\omega}F_{1}(a-c+1,b-c+1,2-c;z).
\end{equation}
In the neighborhood of the horizon $z=0$, the function $\varphi(z)$ behaves as
\begin{equation}
 \varphi(z) = C_{1}e^{-ix_{+}\omega \ln z} + C_{2}e^{ix_{+}\omega \ln z},
\end{equation}
for the scalar field $\phi$ one gets
\begin{equation}
 \phi \sim C_{1}e^{-i\omega \left(t + x_{+}\ln z\right)} + C_{2}e^{-i\omega\left(t - x_{+}\ln z\right)}.
\end{equation}
The first term in the last equation corresponds to an ingoing wave at the BH, while the second one represents an outgoing wave. 
In order to compute the QNMs, we must impose that there exist only ingoing waves at the horizon of the BH, then 
$C_{2} = 0$. The radial solution at the horizon is given by
\begin{equation}
\varphi(z) = C_{1}z^{-ix_{+}\omega}(1-z)^{-ix_{+}\omega}F_{1}(a,b,c;z).
\end{equation}
In order to implement the boundary conditions at infinity, $z=1$, we use the linear transformation $z\rightarrow 1-z$, and then we apply
Kummer's formula \cite{Abramowitz} for the hypergeometric function, 
\begin{eqnarray}
\varphi(z) &=& C_{1}z^{-ix_{+}\omega}(1-z)^{-ix_{+}\omega}\frac{\Gamma(c)\Gamma(c-a-b)}{\Gamma(c-a)\Gamma(c-b)}F_{1}(a,b;a+b-c+1;1-z)\nonumber \\
&+& C_{1}z^{-ix_{+}\omega}(1-z)^{ix_{+}\omega}\frac{\Gamma(c)\Gamma(a+b-c)}{\Gamma(a)\Gamma(b)}F_{1}(c-a,c-b;c-a-b+1;1-z),
\end{eqnarray}
This solution near the infinity, $z=1$, takes the form
\begin{equation}
 \varphi(z) = C_{1}(1-z)^{-ix_{+}\omega}\frac{\Gamma(c)\Gamma(c-a-b)}{\Gamma(c-a)\Gamma(c-b)}
 + C_{1}(1-z)^{ix_{+}\omega}\frac{\Gamma(c)\Gamma(a+b-c)}{\Gamma(a)\Gamma(b)},
\end{equation}
and the scalar field solution near infinity behaves as
\begin{equation}
\phi \sim C_{1}e^{-i\omega(t+x_{+}\ln(1-z))}\frac{\Gamma(c)\Gamma(c-a-b)}{\Gamma(c-a)\Gamma(c-b)}
 + C_{1}e^{-i\omega(t-x_{+}\ln(1-z))}\frac{\Gamma(c)\Gamma(a+b-c)}{\Gamma(a)\Gamma(b)}.
\end{equation}
To compute the QNMs, we also need to impose the boundary conditions on the solution of the radial equation at infinity, meaning that
only purely outgoing waves are allowed there. Therefore, the second term in the last equation must vanish; this is fulfilled, at the 
poles of $\Gamma(a)$ or $\Gamma(b)$, where the scalar field satisfies the considered boundary condition only when
\begin{equation}
 a = -n \ \ \ \ \ \mbox{or} \ \ \ \ \ b=-n,
\end{equation}
where $n = 0, 1, 2,...$. These conditions determine the form of the quasinormal modes,
\begin{equation}
\omega = -\frac{i}{2x_{+}}\left(n + 1\right). 
\end{equation}

\subsection{Spacetime metric  $ds^{2} = \left(\left(\frac{\Lambda}{\eta}\right)x^{2} + 2Mx - \epsilon \right)dt^{2} + 
\frac{1}{\left(\frac{\Lambda}{\eta}\right)x^{2} + 2Mx - \epsilon}dx^{2}$}

For the second metric given in Eq. (\ref{eq:sol3}), we have a spacetime that is not asymptotically flat; as such, and as mentioned before, we use a 
definition for the QNMs different from the one used in an asymptotically flat spacetime. 
The formal treatment for this kind of spacetime is discussed in \cite{Horowitz:1999jd}, where they defined QNMs to be modes with only 
ingoing waves near the horizon and vanishing at infinity. Thus, the Klein-Gordon equation (\ref{eq:eom}) can be written as
\begin{equation}
-\frac{1}{u^{2}-u_{+}^{2}}\partial_{t}\partial_{t}\phi + \frac{2}{l^{2}}u \partial_{u}\phi + 
\frac{1}{l^{2}}(u^{2}-u_{+}^{2})\partial_{u}\partial_{u}\phi - m^{2}\phi=0,
\end{equation}
now, we will consider a solution of type $\phi(t,u) = \varphi(u)e^{-i\omega t}$ and definition $l = (\lambda/\eta)^{1/4}$, for which 
the radial equation can be written as follows
\begin{equation}
\frac{1}{l^{2}}(u^{2}-u_{+}^{2})\varphi''(u) + \frac{2}{l^{2}}u\varphi'(u) 
+ \left(\frac{\omega^{2}}{u^{2}-u_{+}^{2}} - m^{2} \right)\varphi(u) = 0,
\label{eq:radial}
\end{equation}
where the prime denotes derivates with respect the variable $u$. If we define the change of variable $z=1-u_{+}^{2}/u^{2}$ 
\cite{Aros} and follow the procedure stated for the massless case, the equation (\ref{eq:radial}) 
transforms into the hypergeometric differential equation (\ref{eq:hyper}) for the function $F(z)$, where the coefficients $a,b,c$ 
are given by the following relations 
\begin{eqnarray}
a + b &=& 2\alpha + 2\beta + \frac{1}{2},\\
ab &=& \alpha(\alpha-1)+\beta(\beta-1)+2\alpha \beta + \frac{3}{2}\alpha + \frac{3}{2}\beta,\\
c &=& 2\alpha + 1,
\end{eqnarray}
which gives
\begin{eqnarray}
a &=& \alpha + \beta,\\
b &=& \alpha + \beta + \frac{1}{2},  
\end{eqnarray}
and for the exponents $\alpha$ and $\beta$
\begin{eqnarray}
\alpha &=& -i\frac{l}{u_{+}}\omega,\\
\beta &=& \frac{1}{4}\left(1 - \sqrt{1+4m^{2}l^{2}} \right),
\end{eqnarray}
where, without loss of generality, we have chosen the negative signs. The solution of the radial equation reads
\begin{equation}
 F(z) = C_{1}F_{1}(a,b,c;z)+C_{2}z^{1-c}F_{1}(a-c+1,b-c+1,2-c;z),
\end{equation}
where $C_{1}$ and $C_{2}$ are arbitrary constants and $F_{1}(a,b,c;z)$ is the hypergeometric function. Since $\varphi(z) = 
z^{\alpha}(1-z)^{\beta}F(z)$, the behaviour of the scalar field near the horizon ($z=0$) is given by
\begin{equation}
 \phi \sim C_{1}e^{-i\omega \left(t + \frac{l}{u_{+}}\ln z\right)} + C_{2}e^{-i\omega\left(t - \frac{l}{u_{+}}\ln z\right)}.
\end{equation}
Then, the scalar field $\phi$ is purely ingoing at the horizon for $C_{2} = 0$, and therefore the radial solution is 
\begin{equation}
\varphi(z) = C_{1}z^{\alpha}(1-z)^{\beta}F_{1}(a,b,c;z).
\end{equation}

\begin{figure}[h]
\begin{center}
\includegraphics[width=0.6\textwidth]{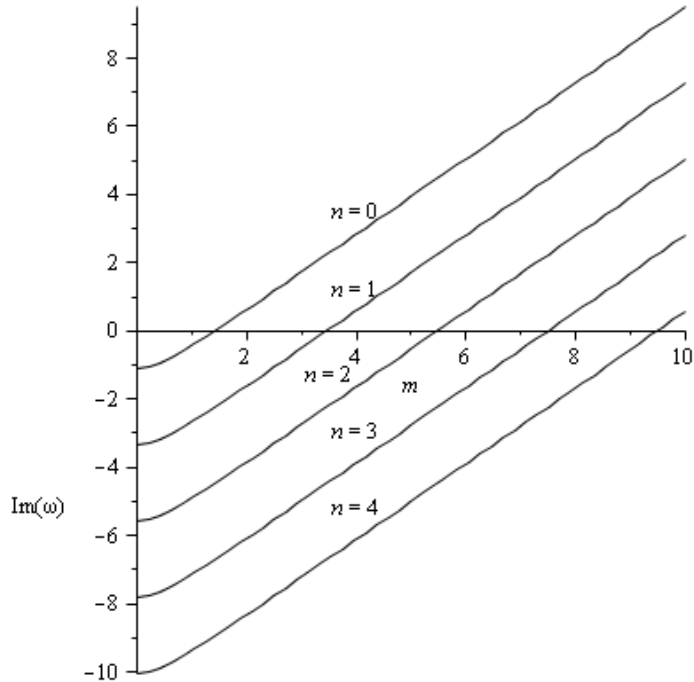}
\end{center}
\caption{In this plot we depict the behaviour of the QNMs expressed in Eq. (\ref{eq:of}), with some paremeters values, $l=1$ and 
$u_{+} = \sqrt{5}$. We can see that the BH becomes unstable for large values of the mass $m$.} 
\label{fig:Pot2}
\end{figure}

In order to implement boundary conditions at infinity ($z=1$), we use the linear transformation $z \rightarrow 1-z$ for the 
hypergeometric function and we obtain
\begin{eqnarray}
\varphi(z) &=& C_{1}z^{\alpha}(1-z)^{\beta}\frac{\Gamma(c)\Gamma(c-a-b)}{\Gamma(c-a)\Gamma(c-b)}F_{1}(a,b;a+b-c+1;1-z)\nonumber \\
&+& C_{1}z^{\alpha}(1-z)^{c-a-b+\beta}\frac{\Gamma(c)\Gamma(a+b-c)}{\Gamma(a)\Gamma(b)}F_{1}(c-a,c-b;c-a-b+1;1-z).
\end{eqnarray}
Using the condition of the flux
\begin{eqnarray}
\mathcal{F} &\sim& \varphi^{*}(z)\partial_{z}\varphi(z)-\varphi(z)\partial_{z}\varphi^{*}(z),\nonumber \\ 
&\sim& -2\,i{C}_{1}^{2}\frac{l\omega}{u_{+}}\, \left( 
 \left( 1-z \right) ^{1/2\left(1-\,\sqrt {1+4\,{m}^{2}{l}^{2}}\right)}{ \Gamma_{1}}
^{2} + 2\, \Gamma_{1}\, \Gamma_{2}\sqrt{1-z}+ \left( 1-z \right) ^{1/2\left(1+\,\sqrt {1
+4\,{m}^{2}{l}^{2}}\right)}{\Gamma_{2}}^{2} \right),\nonumber \\
&\sim& -2\,i{C}_{1}^{2}\frac{l\omega}{u_{+}}\, \left( 
 \left( 1-z \right) ^{2\beta}{ \Gamma_{1}}
^{2} + 2\, \Gamma_{1}\, \Gamma_{2}\sqrt{1-z}+ \left( 1-z \right) ^{1-2\beta}{\Gamma_{2}}^{2} \right),
\label{eq:flux} 
\end{eqnarray}
where
\begin{eqnarray}
 \Gamma_{1} &=& \frac{\Gamma(c)\Gamma(c-a-b)}{\Gamma(c-a)\Gamma(c-b)},\\
\Gamma_{2} &=& \frac{\Gamma(c)\Gamma(a+b-c)}{\Gamma(a)\Gamma(b)},
\end{eqnarray}
then, the flux (\ref{eq:flux}) has a leading term $(1-z)^{1-2\beta}$ and vanishes at infinity only if we impose that
\begin{equation}
a = -n \ \ \ \ \ \mbox{or} \ \ \ \ \ b = -n, 
\end{equation}
where $n = 0, 1, 2, 3...$. These conditions lead to the determination of the quasinormal modes as follows
\begin{equation}
\omega = -i\frac{u_{+}}{4l}\left[4n+3-\sqrt{1+4m^{2}l^{2}}\right].
\label{eq:of} 
\end{equation}
Our results are represented in Fig. (\ref{fig:Pot2}), where it is possible to see that, for a scalar field with large mass, the BH 
becomes unstable, while for lower values of the mass, or $m=0$, this kind of black hole is stable.


\section{Final remarks}
This article was devoted to studying the response of two $1+1$ BH under scalar perturbations. We focused on the BH solutions found in 
Ref. \cite{HL} in the context of HL gravity. The BH studied in the present paper also correspond to solutions arising from 
standard GR plus the dilaton field; therefore, the physical properties of these BH can be used in different contexts. We noted that 
in studying the QNM oscillations of the metric (\ref{eq:lowest}) with massive scalar field perturbations it is necessary to look at the 
solution in terms of the confluent hypergeometric, or Kummer, functions, and, as a result, found two different cases, in one case we have
absent QNMs under scalar perturbations and in the second case we have a discrete spectrum. These results are different from those 
obtained  in Ref. \cite{Lopez}, where the QNMs are a continuous spectrum. Also, we computed the frequencies of the massless scalar field as sources of perturbations, 
and again obtained a discrete spectrum. From these results, we conclude that this BH is stable under massive and massless scalar 
perturbations. 

On the other hand, for spacetime in which the cosmological constant does not vanish, we found, in addition to the exact quasinormal 
frequencies, that it is possible to see that when the mass of the scalar field is large, the geometry becomes unstable.
Finally, we would like to note that the frequencies found in this article are purely imaginary, and as such represent pure damping 
behaviour.


\section*{Acknowledgments}
The authors would like to thank Samuel Lepe and Olivera Miskovic for useful 
comments. MC was supported by PUCV through {\it Proyecto DI Postdoctorado} 2015. 
MG-E acknowledges support from a PUCV doctoral scholarship. 
 



\end{document}